\documentclass[a4paper, 11pt]{article}

\usepackage{jheppub}
\makeatletter
\gdef\@fpheader{}
\makeatother

\usepackage{amsmath}
\usepackage{amssymb}
\usepackage{bm}
\usepackage{mathtools}

\usepackage{graphicx}
\usepackage{booktabs}

\title{\boldmath
On magnetic-field consistency in the sideways force calculation of Phys. Plasmas 27, 012508 (2020)
}

\author[a]{Denis Mironov}

\affiliation[a]{
    Independent Researcher, Mountain Brook, AL, USA
}

\abstract{
We examine the consistency of the equilibrium magnetic field $\mathbf{B}_0$ used in the analytical calculation of the sideways wall force by A. A. Martynov and S. Yu. Medvedev [\href{https://doi.org/10.1063/1.5133960}{Phys. Plasmas \textbf{27}, 012508 (2020)}] and the subsequent reformulation presented in [\href{https://doi.org/10.1063/5.0006627}{Phys. Plasmas \textbf{27}, 054702 (2020)}]. In these attempts, the current density $\mathbf{j}_0 = \nabla\times\mathbf{B}_0$ must vanish in the vacuum regions adjacent to the wall, making $\mathbf{B}_0$ curl-free as a complement to the natural $\nabla\cdot\mathbf{B}_0=0$. Since the cited works incorporate selected toroidal corrections without explicitly specifying the corresponding differential operators, we test $\mathbf{B}_0$ adopted in these models against Maxwell's equations using both toroidal and quasi-cylindrical formulations. We show that the divergence-free requirement is violated in the toroidal interpretation, while the curl-free condition is not satisfied in the quasi-cylindrical limit. The reformulation likewise fails to define a self-consistent equilibrium field with toroidal effects because the accompanying field corrections required for consistency with Maxwell's equations are not specified. Since the sideways force calculations are built upon these inconsistent field prescriptions, their analytical derivations are consequently compromised. The resulting arguments therefore cannot serve as a valid basis for revising either the earlier results established by D. V. Mironov and V. D. Pustovitov [\href{https://doi.org/10.1063/1.4990009}{Phys. Plasmas \textbf{24}, 092508 (2017)}] or those obtained within the single-harmonic analysis [\href{https://doi.org/10.1063/1.4921660}{Phys. Plasmas \textbf{22}, 052502 (2015)}].
}

\begin{document}

\maketitle

\flushbottom



\section{Introduction}
\label{sec:introduction}
In~\cite{Martynov2020}, the authors challenge the analytical expression for the sideways wall force derived in~\cite{Mironov2017} and, on this basis, claim that some conclusions of that model are incorrect. They further argue that the results of the earlier single-harmonic analysis~\cite{Mironov2015} must likewise be re-interpreted.

The analytical part of the work~\cite{Martynov2020} was subsequently discussed in a Comment~\cite{Comment-Pustovitov2020} and in a Response~\cite{Response-Martynov2020}. The Comment~\cite{Comment-Pustovitov2020} primarily addresses the Resistive Wall Mode (RWM) problem, but also briefly questions the justification for the choice of the toroidal component $B_{0\zeta}^{\mathrm{MM}}$ of the ``equilibrium magnetic field" $\mathbf{B}_{0}^{\mathrm{MM}}$ adopted in \cite{Martynov2020}. This selective modification of the physically admissible equilibrium magnetic field $\mathbf{B}_0$ is essential to the ``derivation"~\cite{Martynov2020} of the sideways wall force expression that differs from the result obtained in~\cite{Mironov2017}.

Here we complement the analysis~\cite{Comment-Pustovitov2020} by examining in detail the self-consistency of $\mathbf{B}_{0}^{\mathrm{MM}}$ and its modified version~\cite{Response-Martynov2020}. Preliminaries are given in Sec.~\hyperref[sec:preliminaries]{2}. In Sec.~\hyperref[sec:equilibrium-magnetic-field]{3}, we show that $\mathbf{B}_0^{\mathrm{MM}}$ proposed as the equilibrium magnetic field in~\cite{Martynov2020} is not divergence-free in the toroidal interpretation and not curl-free in the quasi-cylindrical formulation. In Sec.~\hyperref[sec:modification-of-the-magnetic-field]{4}, we show that the reformulation~\cite{Response-Martynov2020} likewise fails to specify the accompanying equilibrium field corrections required for a consistent treatment of toroidal effects. The conclusions are summarized in Sec.~\hyperref[sec:conclusion]{5}.


\section{Preliminaries}
\label{sec:preliminaries}
The analytical treatments~\cite{Martynov2020,  Mironov2017, Comment-Pustovitov2020, Response-Martynov2020, Mironov2015} consider a tokamak configuration in which the plasma is separated from a single resistive wall by a vacuum gap. The sideways force acting on the wall along the fixed horizontal direction $X$ is defined by
\begin{equation}
\label{eq:F_X-def}
    F_X^w 
    \equiv
    \int\limits_{\Omega^w}
    (\mathbf{j}\times\mathbf{B})\cdot\mathbf{e}_X\,\mathrm{d}V,
\end{equation}
where $\Omega^w$ denotes the volume of the wall, $\mathbf{j}=\nabla\times\mathbf{B}$ is the current density, $\mathbf{B}$ is the magnetic field subject to $\nabla\cdot\mathbf{B}=0$, and $\mathbf{e}_X$ is the unit vector pointing along the positive $X$ direction.

By using $\mathbf{j}=\nabla\times\mathbf{B}$, the integrand in~\eqref{eq:F_X-def} can be written as
\begin{equation}
\label{eq:jxB-e_X}
    (\mathbf{j}\times\mathbf{B})\cdot\mathbf{e}_X 
    =
    \nabla\cdot
    \left\{
        (\mathbf{B}\cdot\mathbf{e}_X)\,\mathbf{B} - \frac{\mathbf{B}^2}{2}\,\mathbf{e}_X
    \right\}.
\end{equation}
With this and using the well-known divergence theorem, $F_X^w$ in~\eqref{eq:F_X-def} can equivalently be expressed as an integral over the boundary of the wall volume~\cite{Mironov2017, Pustovitov2015, LandauLifshitz-electrodynamics-vol8}
\begin{equation}
\label{eq:F_X=F_X-out-F_X-in} 
   F_{X}^w 
   = 
   F_{X}^{\mathrm{out}} - F_{X}^{\mathrm{in}},
\end{equation}
where
\begin{equation}
\label{eq:F_X-alpha}
    F_{X}^{\alpha} 
    = 
    \oint\limits_{S_\alpha^w}
    \left\{
        (\mathbf{B}\cdot\mathbf{e}_X)\,\mathbf{B} - \frac{\mathbf{B}^2}{2}\,\mathbf{e}_X
    \right\}
    \cdot\mathrm{d}\mathbf{S}_\alpha.
\end{equation}
Here $\mathrm{d}\mathbf{S}_\alpha=\mathbf{n}_\alpha\mathrm{dS}_\alpha$ is the vector surface element with the unit normal $\mathbf{n}_\alpha$ pointing outward from the plasma. The subscript $\alpha=\mathrm{out}$ or $\alpha=\mathrm{in}$ refers to the outer, $S_\mathrm{out}^w$, or inner, $S_\mathrm{in}^w$, surface of the wall, respectively. We use the same units for $\mathbf{B}$ as in~\cite{Mironov2015, Mironov2017, Comment-Pustovitov2020, Martynov2020, Pustovitov2015, Response-Martynov2020}. Then the magnetic pressure is $\mathbf{B}^2/2$ and the conversion factor to the SI system is $1/\mu_0$ with the vacuum magnetic permeability $\mu_0=4\pi \times 10^{-7}$ H/m.

Following the concept of a standard perturbation theory, the total magnetic field $\mathbf{B}$ can be decomposed as
\begin{equation}
\label{eq:B=B0+b}
    \mathbf{B} 
    = 
    \mathbf{B}_0 
    + 
    \mathbf{b}
\end{equation}
with the axisymmetric time-independent equilibrium field $\mathbf{B}_0$ and time-dependent perturbation $\mathbf{b}$. Substituting~\eqref{eq:B=B0+b} into~\eqref{eq:F_X-alpha} gives
\begin{equation}
\label{eq:F_X_i_three-terms}
\begin{gathered}
    F_X^\alpha
    =
    \oint\limits_{S_\alpha^w}
    \Bigl\{
        (\mathbf{e}_X\cdot\mathbf{b})\,\mathbf{B}_0
        +
        (\mathbf{e}_X\cdot\mathbf{B}_0)\,\mathbf{b}
        -
        (\mathbf{b}\cdot\mathbf{B}_0)\,\mathbf{e}_X
    \Bigr\}\cdot\mathrm{d}\mathbf{S}_\alpha
    \\[0.5em]
    + \
    \oint\limits_{S_\alpha^w}
    \Bigl\{
        (\mathbf{b}\cdot\mathbf{e}_X)\,\mathbf{b}
        -
        \frac{\mathbf{b}^2}{2}\,\mathbf{e}_X
    \Bigr\}\cdot\mathrm{d}\mathbf{S}_\alpha
    \\[0.5em]
    + \
    \oint\limits_{S_\alpha^w}
    \Bigl\{
        (\mathbf{B}_0\cdot\mathbf{e}_X)\,\mathbf{B}_0
        -
        \frac{\mathbf{B}_0^2}{2}\,\mathbf{e}_X
    \Bigr\}\cdot\mathrm{d}\mathbf{S}_\alpha.
\end{gathered}
\end{equation}
In~\cite{Mironov2017, Martynov2020}, the sideways wall force was evaluated to linear order in $\mathbf{b}$, while the quadratic terms were neglected under the assumption $\mathbf{b}\ll\mathbf{B}_0$. The third integral in~\eqref{eq:F_X_i_three-terms} is essentially~\eqref{eq:F_X-alpha} with $\mathbf{B}=\mathbf{B}_0$. For configurations discussed in~\cite{Martynov2020, Response-Martynov2020, Mironov2015, Mironov2017, Comment-Pustovitov2020}, $F_X^\alpha(\mathbf{B}_0)=0$.

Under the aforementioned assumptions, the analysis presented in~\cite{Martynov2020, Response-Martynov2020, Mironov2017, Comment-Pustovitov2020} focuses on the first integral in~\eqref{eq:F_X_i_three-terms}:
\begin{equation}
\label{eq:F_X_alpha}
    F_{X}^{\alpha} 
    = 
    \oint\limits_{S_\alpha^w} 
        \{
            (\mathbf{e}_{X}\cdot\mathbf{b})\,\mathbf{B}_{0}
            +
            (\mathbf{e}_{X}\cdot\mathbf{B}_{0})\,\mathbf{b}
            -
            (\mathbf{b}\cdot\mathbf{B}_{0})\,\mathbf{e}_{X}
        \}
        \cdot \mathrm{d}\mathbf{S}_{\alpha}.
\end{equation}
Calculating~\eqref{eq:F_X_alpha} requires $\mathbf{B}_{0}$ and $\mathbf{b}$ to be self-consistent at each step of the derivation. In~\cite{Mironov2015, Mironov2017}, the equilibrium field was specified within the quasi-cylindrical approximation, whereas selected toroidal corrections to this field were introduced in~\cite{Martynov2020, Response-Martynov2020}. Therefore, the field prescriptions~\cite{Martynov2020, Response-Martynov2020} require additional consistency checks before they can be regarded as a self-consistent equilibrium field with toroidal effects for evaluating~\eqref{eq:F_X_alpha}.


\section{Equilibrium magnetic field in \texorpdfstring{Ref~\protect\citenum{Martynov2020}}{Martynov and Medvedev (2020)}}
\label{sec:equilibrium-magnetic-field}

\begin{figure}[!t]
    \centering
    \includegraphics[
        width=0.6\linewidth,
        trim=1.5bp 0bp 185.25bp 0bp,
        clip]{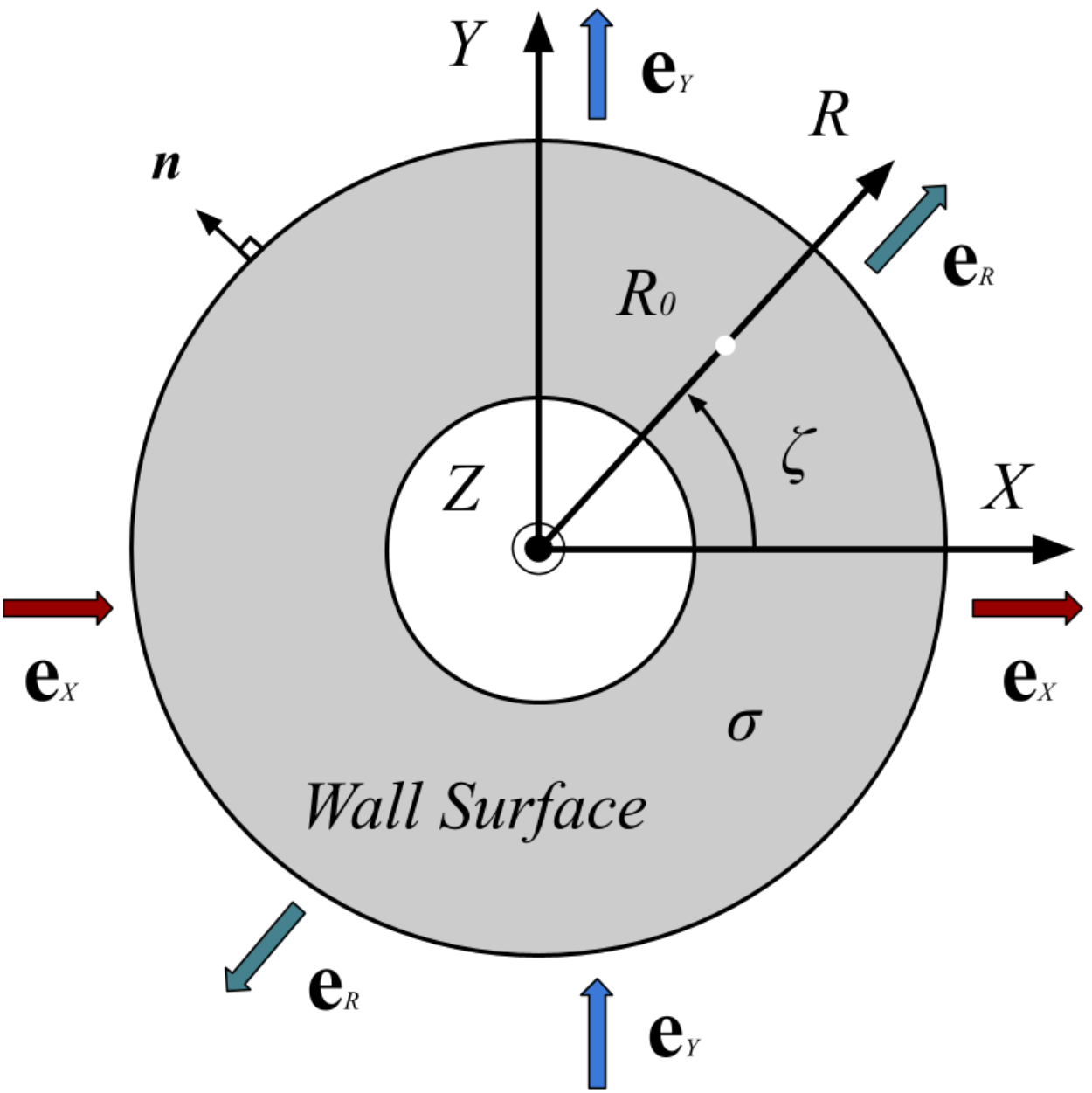}
    \caption{Two right-handed coordinate systems, namely cylindrical $(R,\zeta,Z)$ and Cartesian $(X,Y,Z)$ so that $(X,Y) = (R\cos\zeta, R\sin\zeta)$, are shown from the top of the tokamak toroidal wall with the conductivity $\sigma$ and the centerline located at $R=R_0$. The circled dot in the middle corresponds to the positive $+Z$ direction pointing out of the page. Two perpendicular black coordinate axes represent $+X$ and $+Y$. The ray labeled $R$, lying between them, indicates the cylindrical radial direction at the specified toroidal angle $\zeta$. The latter is measured from the $+X$ axis. The short red and blue arrows indicate the fixed Cartesian unit vectors $\mathbf{e}_X=\nabla X$ and $\mathbf{e}_Y=\nabla Y$, respectively. The teal arrows show the cylindrical radial unit vector $\mathbf{e}_R=\nabla R$ at two opposite toroidal locations, illustrating its variation with $\zeta$. The thin black arrow $\mathbf{n}$ denotes the outward-pointing unit normal to the wall surface.}
    \label{fig:toroidal-cross-section}
\end{figure}

\begin{figure}[!t]
    \centering
    \includegraphics[width=0.7\linewidth]{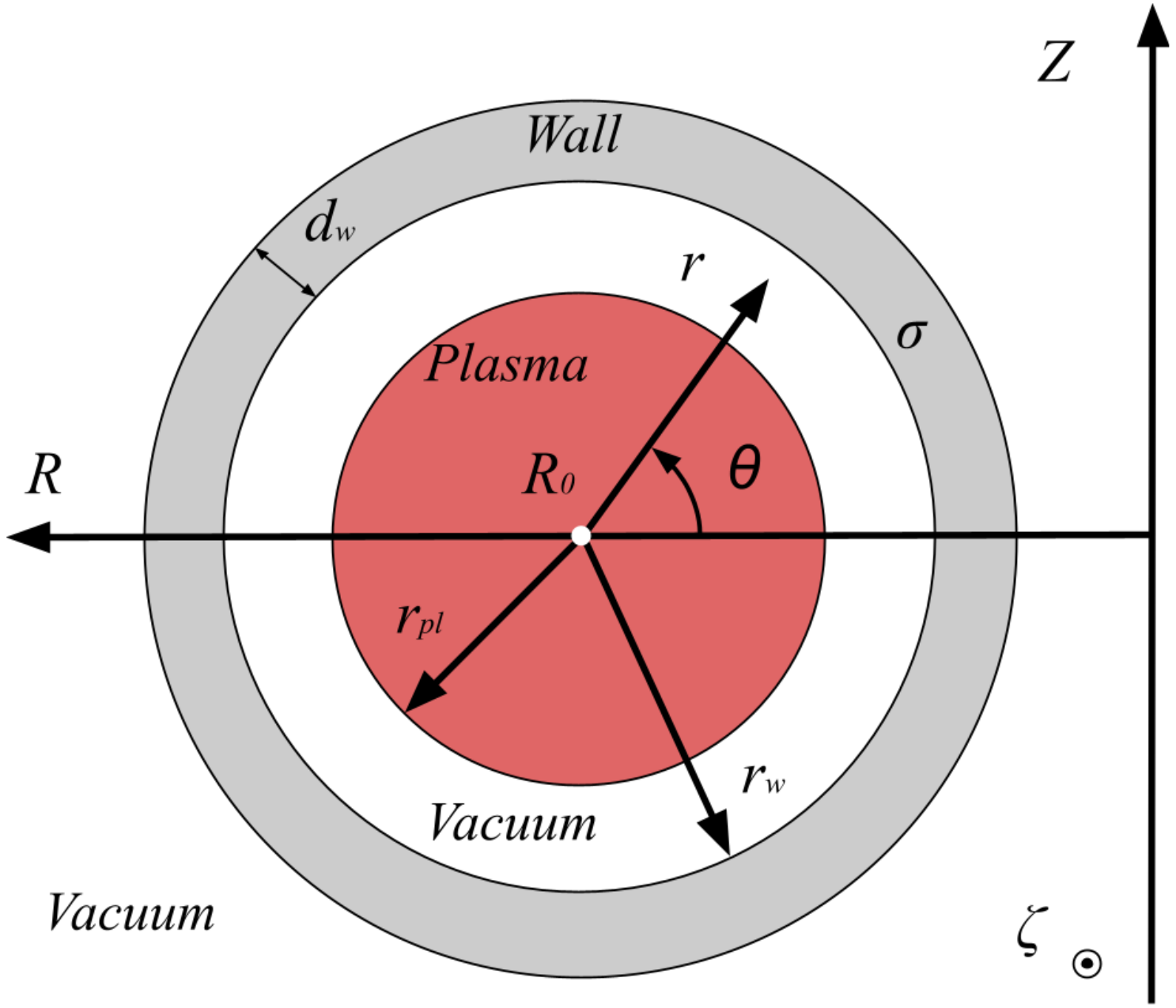}
    \caption{Two right-handed coordinate systems, namely cylindrical $(R,\zeta,Z)$ and quasi-cylindrical $(r,\theta,\zeta)$ so that $(R,Z) = (R_0-r\cos\theta, r\sin\theta)$, are shown in the poloidal cross-section of a circular tokamak at fixed toroidal angle $\zeta$. The positive direction $+\zeta$ points out of the page, as indicated by the circled dot. The poloidal angle $\theta$ is measured from the side nearest to the $Z$ axis, where at $\theta=0$ both $R=R_0-r$ and $Z=0$. The red disk represents the plasma of minor radius $r_{pl}$ and the centerline at $R=R_0$. The gray annulus represents the resistive wall of conductivity $\sigma$, thickness $d_w$, and the inner surface at $r_w$. The plasma--wall gap and the region outside the wall are vacuum.}
    \label{fig:poloidal-cross-section}
\end{figure}

The tokamak geometry and coordinate conventions are illustrated in Figs.~\ref{fig:toroidal-cross-section} and~\ref{fig:poloidal-cross-section}, which schematically show the toroidal top view and the poloidal cross-section, respectively. 

The analysis of $\mathbf{B}_{0}$, entering~\eqref{eq:F_X_alpha}, starts with the general tensor representations
\begin{equation}
{\label{eq:divB_0}}
    \nabla \cdot \mathbf{B}_{0} 
    = 
    \frac{1}{\sqrt{g}}
    \frac{\partial}{\partial q^{i}}\,(\sqrt{g}B_{0}^i)
\end{equation}
and
\begin{equation}
\label{eq:rotB}
    (\nabla \times \mathbf{B}_{0})^{i} 
    = 
    \frac{1}{\sqrt{g}}\,
    \varepsilon^{ijk}\frac{\partial}{\partial q^{j}}\,(g_{kl}B_{0}^{l}),
\end{equation}
where Einstein summation over repeated indices is implied unless stated otherwise. The notations are as follows: 
\begin{equation}
\label{eq:g-def}
    g=\mathrm{det}[g_{ij}]
\end{equation}
with the metric-tensor components $g_{ij}$ defined as
\begin{equation}
\label{eq:g_{ij}-def}
    g_{ij}
    \equiv
    \begin{cases}
    \displaystyle
    \left\|
    \frac{\partial \boldsymbol{\rho}}{\partial q^{i}}
    \right\|_2^2,
    & i=j,
    \\[4.0ex]
    \displaystyle
    \frac{\partial \boldsymbol{\rho}}{\partial q^{i}}
    \cdot
    \frac{\partial \boldsymbol{\rho}}{\partial q^{j}},
    & i\neq j.
    \end{cases}
\end{equation}
\par\noindent Here $\left\|\cdot\right\|_2$ denotes the Euclidean norm induced by the standard inner product. The curvilinear coordinates $q^i$ are specialized to the right-handed quasi-cylindrical coordinate system $(r,\theta,\zeta)$ as
\begin{equation}
    (q^1,q^2,q^3) = (r,\theta,\zeta).
\end{equation}
The position vector $\boldsymbol{\rho}$ is represented in the orthonormal Cartesian basis $\{\mathbf{e}_{X}, \mathbf{e}_{Y}, \mathbf{e}_{Z}\}$ as
\begin{equation}
\label{eq:rho-def}
    \boldsymbol{\rho}
    \equiv
    \begin{pmatrix}
        X \\ Y \\ Z
    \end{pmatrix}
    =
    \begin{pmatrix}
        R\cos\zeta \\
        R\sin\zeta \\
        r\sin\theta \\
    \end{pmatrix},    
\end{equation}
where
\begin{equation}
\label{eq:R-def}
    R \equiv R_0 - r\cos\theta.
\end{equation}
The totally antisymmetric Levi-Civita symbol is $\varepsilon^{ijk}$ with $\varepsilon^{123}=+1$, and $B_{0}^{i}$ are contravariant coordinate components  of $\mathbf{B}_{0}$.

From \eqref{eq:g_{ij}-def}--\eqref{eq:R-def}, 
\begin{equation}
    [g_{ij}] = \operatorname{diag}\!\left(1,\, r^2,\, R^2\right)
\end{equation}
is a diagonal matrix and, as follows from~\eqref{eq:g-def},
\begin{equation}
    \sqrt{g}=rR.
\end{equation}
The contravariant components $B_{0}^{i}$ are expressed in terms of the physical components $B_{0\hat{i}}$ of $\mathbf{B}_{0}$ given in the right-handed orthonormal basis $\{\mathbf{e}_{r}, \mathbf{e}_{\theta}, \mathbf{e}_{\zeta} \}$ as
\begin{equation}
\label{eq:B0r-B0theta-B0zeta}
    B_{0}^{1} = \frac{B_{0r}}{h_{1}}, 
    \quad
    B_{0}^{2} = \frac{B_{0\theta}}{h_{2}},
    \quad
    B_{0}^{3} = \frac{B_{0\zeta}}{h_{3}},
\end{equation}
where the superscript labels the corresponding coordinate component and 
\begin{equation}
\label{eq:Lame-coeff}
    h_{1} = 1,
    \quad
    \quad
    h_{2} = r,
    \quad
    \quad
    h_{3} = R
\end{equation}
being the Lam\'e coefficients that are defined as $h_{i} \equiv \sqrt{g_{ii}}$ with $g_{ii}$ introduced in~\eqref{eq:g_{ij}-def}. 

With these relations, \eqref{eq:divB_0} gives
\begin{equation}
\label{eq:divB0-tor}
    \nabla \cdot \mathbf{B}_{0} 
    = 
    \frac{1}{rR}
    \left[
        \frac{\partial (rRB_{0r})}{\partial r}
        +
        \frac{\partial (RB_{0\theta})}{\partial \theta}
        +
        \frac{\partial (rB_{0\zeta})}{\partial \zeta}
    \right],
\end{equation}
while~\eqref{eq:rotB}, upon transformation to physical components 
\begin{equation}
    (\nabla\times\mathbf{B}_{0})_{\hat{i}} = h_{i} (\nabla\times\mathbf{B}_{0})^{i}
\end{equation}
with no summation over $i$, gives
\begin{equation}
\label{eq:rotB0-r}
    (\nabla \times \mathbf{B}_{0})_{r} 
    = 
    \frac{1}{rR}
    \left[
        \frac{\partial (RB_{0\zeta})}{\partial\theta}
        -
        \frac{\partial (rB_{0\theta})}{\partial\zeta}
    \right],
\end{equation}
\begin{equation}
\label{eq:rotB0-theta}
    (\nabla \times \mathbf{B}_{0})_{\theta} 
    =
    \frac{1}{R}
    \left[
        \frac{\partial B_{0r}}{\partial\zeta}
        -
        \frac{\partial (RB_{0\zeta})}{\partial r}
    \right],
\end{equation}
\begin{equation}
\label{eq:rotB0-zeta}
    (\nabla \times \mathbf{B}_{0})_{\zeta} 
    = 
    \frac{1}{r}
    \left[
        \frac{\partial (rB_{0\theta})}{\partial r}
        -
        \frac{\partial B_{0r}}{\partial \theta}
    \right].
\end{equation}
Equations~\eqref{eq:divB0-tor} and \eqref{eq:rotB0-r}--\eqref{eq:rotB0-zeta} are derived for a circular tokamak considered in the analytical treatments~\cite{Martynov2020, Mironov2017}.

In the analytical model~\cite{Mironov2017}, the calculation of \eqref{eq:F_X_alpha} was performed for
\begin{equation}
\label{eq:B0-MP}
    \mathbf{B}_0 
    = 
    B_{0\theta} \mathbf{e}_{\theta} 
    + 
    B_{0\zeta} \mathbf{e}_{\zeta},
\end{equation}
where $rB_{0\theta}$ and $B_{0\zeta}$ were treated as constants outside the plasma. In~\cite{Martynov2020}, the only specified modification to this equilibrium field is the introduction of $R_0/R$ dependence into its toroidal component at the wall, without corresponding corrections to the other components. Taking this prescription as the field expression to be tested gives
\begin{equation}
\label{eq:B0-MM}
    \mathbf{B}_{0}^{\mathrm{MM}} 
    \equiv 
    B_{0\theta} \mathbf{e}_{\theta} 
    +
    B_{0\zeta}^{\mathrm{MM}} \mathbf{e}_{\zeta}
\end{equation} 
with
\begin{equation}
\label{eq:B0zeta-correction}
    B_{0\zeta}^{\mathrm{MM}}
    \equiv
    \frac{R_{0}B_{0\zeta}}{R_{0} - r \cos\theta}.
\end{equation}
The introduction of this toroidal correction in~\cite{Martynov2020} was subsequently described in~\cite{Response-Martynov2020} as follows: ``The remedy proposed in~\cite{Martynov2020} was to take into account the toroidal correction for the toroidal magnetic field $B_{0\zeta} \rightarrow B_{0\zeta}R_{0}/R$, leading to a consistent sideways force expression for a single harmonic mode in the limit of large aspect ratio." This modification is essential to the revised sideways force expression obtained in~\cite{Martynov2020}, which differs from the result of~\cite{Mironov2017}. It is clear that the prescriptions~\eqref{eq:B0-MM}--\eqref{eq:B0zeta-correction} artificially combine the quasi-cylindrical treatment of $B_{0\theta}$ with the toroidal $1/R$ dependence imposed on $B_{0\zeta}$. It is therefore necessary to examine whether the resulting field satisfies Maxwell's equations in the vacuum regions adjacent to the inner, $S_\mathrm{in}^w$, and outer, $S_\mathrm{out}^w$, wall surfaces entering~\eqref{eq:F_X_alpha}, namely
\begin{equation}
\label{eq:Maxwell-vac}
    \nabla\cdot\mathbf{B}_{0} = 0
    \quad
    \quad
    \mathrm{and}
    \quad
    \quad
    \nabla\times\mathbf{B}_0=\mathbf{0}.
\end{equation}

It remains unclear whether differential operators $\nabla \cdot$ and $\nabla \times$ are to be understood in the toroidal or quasi-cylindrical sense~\cite{Martynov2020, Response-Martynov2020}. These two interpretations are examined below.

We start with the toroidal formulation. Accounting for axisymmetry $(\partial/\partial\zeta=0)$ and substituting $\mathbf{B}_{0}^{\mathrm{MM}}$ for $\mathbf{B}_{0}$ into \eqref{eq:divB0-tor} and \eqref{eq:rotB0-r}--\eqref{eq:rotB0-zeta}, we obtain
\begin{equation}
\label{eq:div-B0-MM-ne-0}
    \nabla\cdot\mathbf{B}_{0}^{\mathrm{MM}}
    =
    \cfrac{B_{0\theta}\sin\theta}{R_0 - r\cos\theta}
    \qquad
    \text{and}
    \qquad
    \nabla\times\mathbf{B}_{0}^{\mathrm{MM}}
    =
    \mathbf{0},
\end{equation}
respectively. Thus, $\mathbf{B}_{0}^{\mathrm{MM}}$ is \textit{not} divergence-free in the toroidal interpretation. In contrast, if the quasi-cylindrical formulation is intended, we have
\begin{equation}
\label{eq:max-second-pair}
    \nabla_{\mathrm{q}} \cdot \mathbf{B}_{0}^{\mathrm{MM}} = 0
    \qquad
    \text{and}
    \qquad
    \nabla_{\mathrm{q}} \times \mathbf{B}_{0}^{\mathrm{MM}}
    = -\cfrac{R_{0} B_{0\zeta}}{(R_0 - r\cos\theta)^2}
    \left(
        \mathbf{e}_{r}\sin\theta
        + \mathbf{e}_{\theta}\cos\theta
    \right).
\end{equation}
Here $\nabla_{\mathrm{q}}\cdot$ and $\nabla_{\mathrm{q}}\times$ denote the quasi-cylindrical limit of $\nabla\cdot$ and $\nabla\times$, respectively, obtained by replacing $R$ with $R_0$, while retaining the exact $B_{0\zeta}^{\mathrm{MM}}$ given by \eqref{eq:B0zeta-correction}. Hence, $\mathbf{B}_{0}^{\mathrm{MM}}$ is $not$ curl-free when $\nabla_{\mathrm{q}}$ is used while the $R$-correction suggested in~\cite{Martynov2020} is retained. 

Equations~\eqref{eq:div-B0-MM-ne-0}--\eqref{eq:max-second-pair} show that $\mathbf{B}_0^{\mathrm{MM}}$ violates Maxwell's equations~\eqref{eq:Maxwell-vac}. The asymptotic orders of these residuals can be identified by expanding the corresponding expressions in the inverse aspect ratio. 

Indeed, introducing a large-aspect ratio parameter
\begin{equation}
\label{eq:epsilon}
    \epsilon\equiv r/R_0
\end{equation}
and using the expansion
\begin{equation}
    \frac{R_0}{R_0 - r\cos\theta} 
    = 
    1 + \epsilon\cos\theta + O(\epsilon^2),    
\end{equation}
the first equation in~\eqref{eq:div-B0-MM-ne-0} gives
\begin{equation}
\label{eq:divergence-toroidal}
    \nabla\cdot\mathbf{B}_0^{\mathrm{MM}}
    =
    \frac{B_{0\theta}}{r}\left[\epsilon\sin\theta + O(\epsilon^2)\right],
\end{equation}
while the second in~\eqref{eq:max-second-pair} yields
\begin{equation}
\label{eq:curl-quasi-cylinder}
    \nabla_\mathrm{q}\times\mathbf{B}_0^\mathrm{MM}
    =
    -\frac{B_{0\zeta}}{r} 
        \left[
            \mathbf{e}_r\epsilon\sin\theta + \mathbf{e}_\theta\epsilon\cos\theta
            +
            O(\epsilon^2)
        \right]. 
\end{equation}
The residuals~\eqref{eq:divergence-toroidal} and~\eqref{eq:curl-quasi-cylinder} are both of first order in $\epsilon$ when normalized by $B_{0\theta}/r$ and $B_{0\zeta}/r$, respectively. Their implications for the retained order of the sideways force calculation depend on the relative ordering of the magnetic field components and their contributions to the force integral.

The analysis demonstrates that the field prescription~\eqref{eq:B0-MM}--\eqref{eq:B0zeta-correction}, when interpreted as a complete equilibrium magnetic field in the toroidal vacuum geometry, does not satisfy the divergence-free condition. Moreover, retaining the toroidal correction to $B_{0\zeta}$ while employing the quasi-cylindrical differential operators leads to the nonzero curl in~\eqref{eq:max-second-pair}, reflecting an inconsistent treatment of terms of the same geometrical order. Thus, the prescription adopted in~\cite{Martynov2020} does not, by itself, define a Maxwell-consistent toroidal equilibrium field.


\section{Modification proposed in \texorpdfstring{Ref~\protect\citenum{Response-Martynov2020}}{Martynov and Medvedev (2020)}}
\label{sec:modification-of-the-magnetic-field}

In~\cite{Response-Martynov2020}, the authors agree with the criticism~\cite{Comment-Pustovitov2020} concerning the toroidal correction applied to $B_{0\zeta}$ in~\cite{Martynov2020}. They explicitly say, ``We agree with the criticism in the Comment concerning this $ad$ $hoc$ choice", and subsequently add, ``More thorough analysis of the mode structure shows that the necessary $R_0/R$ multiplier comes from the perturbed radial 
magnetic field $b_{r}$, generated by the plasma boundary displacement $\xi_{r}$, $b_{r}=i\frac{B_{\theta}}{r}(m-nq)\xi_{r}$, where $B_{\theta} = \frac{R_0}{R}B_{0\theta}$ is the poloidal field with toroidal correction." Nevertheless, the authors~\cite{Response-Martynov2020} still claim the validity of their analytical expression for the force~\cite{Martynov2020}.

Regardless of how such changes are incorporated, $\mathbf{B}_0$ must be divergence-free and, in the region with $\mathbf{j}_0=\mathbf{0}$, curl-free at the same time. With the first requirement $\left(\nabla \cdot \mathbf{B}_0 = 0\right)$ and axisymmetry $\left(\partial / \partial \zeta = 0\right)$, Eq.~\eqref{eq:divB0-tor} gives
\begin{equation}
\label{eq:divB0-SecII}
    \cfrac{\partial (rRB_{0r})}{\partial r}
    + 
    \cfrac{\partial (RB_{0\theta})}{\partial \theta}
    = 0,
\end{equation}
while Eq. \eqref{eq:rotB0-zeta} with $(\nabla \times \mathbf{B}_{0})_{\zeta} = 0$ yields
\begin{equation}
\label{eq:rotB0-zeta-SecII}
    \cfrac{\partial (rB_{0\theta})}{\partial r}
    -
    \cfrac{\partial B_{0r}}{\partial\theta}
    =
    0.
\end{equation}
If $\mathbf{B}_0$ is restricted to the poloidal and toroidal components retained in~\cite{Martynov2020, Response-Martynov2020}, Eq.~\eqref{eq:divB0-SecII} and~\eqref{eq:rotB0-zeta-SecII} require
\begin{equation}
    RB_{0\theta} = g_{1}(r)
    \quad
    \text{and}
    \quad
    rB_{0\theta} = g_{2}(\theta),
\end{equation}
where $g_{1}(r)$ and $g_{2}(\theta)$ are functions of $r$ and $\theta$, respectively. Combined, this system of two equations reduces to 
\begin{equation}
\label{eq:g1-g2-condition}
    rg_{1}(r) = (R_0 - r\cos\theta)g_{2}(\theta).
\end{equation}
For this to be valid for any permissible $r$ and $\theta$, $g_{2}$ must vanish identically. 

Indeed, taking a derivative with respect to $\theta$, the expression \eqref{eq:g1-g2-condition} gives
\begin{equation}
\label{eq:g'-g-connection}
    R_0g'_{2}(\theta) + r[g_2(\theta)\sin\theta - g'_{2}(\theta)\cos\theta] = 0,    
\end{equation}
which, for arbitrary $r$, requires $g_{2}'(\theta)=0$. The latter with \eqref{eq:g'-g-connection} leads to $g_2(\theta)\sin\theta = 0$ and, hence, $g_{2}(\theta) = 0$, which is equivalent to $B_{0\theta} = 0$. Thus, a nonvanishing $B_{0\theta}$ cannot coexist with $B_{0r} \equiv 0$ throughout $\mathbf{j}_0=\mathbf{0}$ region when toroidal effects in $\mathbf{B}_0$ are incorporated.

We can conclude that introducing a toroidal correction exclusively to the poloidal field, while keeping $B_{0r} = 0$, does not define a self-consistent equilibrium magnetic field in the physical system. The necessity of retaining a generally nonzero $B_{0r}$ arises precisely when the corresponding toroidal geometry is properly and carefully accounted for at each step of the derivation. This was not done in the analytical treatments~\cite{Martynov2020, Response-Martynov2020}.

More importantly, if $B_{\theta}=B_{0\theta}R_0/R$ proposed in~\cite{Response-Martynov2020} is substituted for $B_{0\theta}$ in Eq.~\eqref{eq:rotB0-zeta-SecII}, the latter requires 
\begin{equation}
\label{eq:B0r-asimpt}
\frac{\partial B_{0r}}{\partial\theta} = O\left(\epsilon B_{0\theta} \right),   
\end{equation}
where $\epsilon$ is a large-aspect ratio parameter defined by~\eqref{eq:epsilon}. Equation~\eqref{eq:B0r-asimpt} shows that retaining toroidal corrections to the poloidal equilibrium field generally requires an accompanying radial component. The prescription $B_\theta=B_{0\theta}R_0/R$, if interpreted as a vacuum equilibrium field with $rB_{0\theta}=const$, cannot satisfy both Maxwell equations merely by introducing $B_{0r}$ while leaving $B_\theta$ unchanged. Indeed, substituting $B_{\theta}$ for $B_{0\theta}$ in Eq.~\eqref{eq:rotB0-zeta-SecII} gives $\partial B_{0r} / \partial\theta = rB_{0\theta}R_0\cos\theta / R^2$, demonstrating that the $\theta$-dependent radial component is of order $\epsilon B_{0\theta}$. Thus, the modification described in~\cite{Response-Martynov2020} does not, by itself, specify a self-consistent toroidal equilibrium magnetic field.

Now we briefly show how~\eqref{eq:F_X_alpha} changes in a circular tokamak with $\mathrm{d}\mathbf{S}_\alpha=\mathbf{e}_r\mathrm{dS}_\alpha$ when $B_{0r} \ne 0$. With these and a substitution (see~$(15)$--$(16)$ in~\cite{Mironov2017})
\begin{equation}
    \mathbf{e}_X
    =
    \mathbf{e}_\theta\sin\theta\cos\zeta
    -
    \mathbf{e}_r\cos\theta\cos\zeta
    -
    \mathbf{e}_\zeta\sin\zeta    
\end{equation}
into~\eqref{eq:F_X_alpha}, the latter gives
\begin{equation}
\label{eq:F_X-w-expanded}
\begin{gathered}
    F_X^\alpha 
    =
    \oint\limits_{S_\alpha^w}
    \Bigl[
        B_{0\theta}b_r\sin\theta\cos\zeta - B_{0\zeta}b_r\sin\zeta 
        +
        (B_{0\theta}b_\theta + B_{0\zeta}b_\zeta)\cos\theta\cos\zeta
    \Bigr]
    \mathrm{dS}_\alpha
    \\[0.5em]
    + \
    \oint\limits_{S_\alpha^w} 
        (\mathbf{e}_X\cdot\mathbf{b})B_{0r}
    \mathrm{dS}_\alpha.
\end{gathered}
\end{equation}
This expression demonstrates that $B_{0r}\ne 0$ introduces an additional contribution that is absent from the field representation employed in the analytical treatments~\cite{Martynov2020, Response-Martynov2020}. Its magnitude depends on the retained asymptotic order and the mode structure of the magnetic perturbation $\mathbf{b}$. Consequently, a consistent toroidal calculation requires the equilibrium field and its corrections to be determined before the corresponding force contributions can be assessed.

Let us note that the self-consistent quasi-cylindrical limit with $R=R_0$ is free from the requirement~\eqref{eq:B0r-asimpt}. In this case, outside the plasma, $rB_{0\theta}=\mathrm{const}$ and $B_{0\zeta}=\mathrm{const}$. With $B_{0r}=0$, this is precisely the choice of $\mathbf{B}_{0}$ in~\cite{Mironov2017}. With this $\mathbf{B}_0$ and the surface element $\mathrm{dS}_{out} \cong \mathrm{dS}_{in} = r_wR_0\mathrm{d}\theta\mathrm{d}\zeta$, the expression~\eqref{eq:F_X-w-expanded} reduces to Eq.~$(18)$ from~\cite{Mironov2017}.


\section{Conclusion}
\label{sec:conclusion}
The selective toroidal corrections applied to the equilibrium magnetic field in~\cite{Martynov2020} and subsequently reinterpreted in~\cite{Response-Martynov2020} do not constitute a self-consistent treatment of the equilibrium magnetic field. In~\cite{Martynov2020}, the modified $\mathbf{B}_{0}$ is not divergence-free when evaluated using toroidal differential operators and is not curl-free when its toroidal variation is combined with quasi-cylindrical differential operators. The reformulation proposed in~\cite{Response-Martynov2020} likewise does not define a self-consistent equilibrium field with toroidal corrections, since the accompanying corrections to $\mathbf{B}_0$ required by Maxwell's equations are not specified. Since the introduced toroidal corrections are essential to the revised sideways force calculation, the resulting expression lacks a demonstrated self-consistent electromagnetic derivation. These results, combined with the issues identified earlier in~\cite{Comment-Pustovitov2020}, undermine the mathematical and physical justification of the analytical arguments presented in~\cite{Martynov2020, Response-Martynov2020}. Therefore, the calculations in~\cite{Martynov2020, Response-Martynov2020}, as presented, cannot serve as a valid basis for revising either the results established in~\cite{Mironov2017} or those obtained within the single-harmonic analysis~\cite{Mironov2015}.

\bibliographystyle{aipnum4-2}
\bibliography{refs}

\end{document}